%% file: main.tex
\documentclass[%
 reprint,
 amsmath,amssymb,
 aps,
floatfix,
twocolumn,
]{revtex4-1}

\usepackage[utf8]{inputenc}
\usepackage[euler-digits,euler-hat-accent]{eulervm}
\usepackage{newpxtext}

\usepackage{subfig}
\usepackage{subfloat}
\usepackage{amsthm, thmtools}
\usepackage{booktabs}
\usepackage{xfrac}
\usepackage{algorithm}
\usepackage{hyperref}
\usepackage{nameref, cleveref}
\crefname{ineq}{inequality}{inequalities}
\Crefname{ineq}{Inequality}{Inequalities}
\crefname{figure}{Fig.}{Figs.}
\usepackage{graphicx}
\usepackage{epstopdf}

\usepackage{thmtools}
\usepackage{thm-restate}
\usepackage{tikz}
\usetikzlibrary{positioning}
\definecolor{processblue}{cmyk}{0.96,0,0,0}

\graphicspath{{.}}
\usepackage{todonotes}
\usepackage{mathtools}
\DeclarePairedDelimiter\bra{\langle}{\rvert}
\DeclarePairedDelimiter\ket{\lvert}{\rangle}
\DeclarePairedDelimiterX\braket[2]{\langle}{\rangle}{#1 \delimsize\vert #2}
\DeclarePairedDelimiter\set{\{}{\}}

\DeclarePairedDelimiter\paren{(}{)}

\usepackage{algpseudocode}
\usepackage{pgfplots}
\usetikzlibrary{quantikz}

\declaretheorem[name=Theorem, refname={Theorem,Theorems}, Refname={Theorem, Theorems}]{thm}
\declaretheorem[name=Lemma, refname={Lemma, Lemmas}, Refname={Lemma, Lemmas}, sibling=thm]{lem}
\declaretheorem[name=Proposition, refname={Proposition, Propositions}, Refname={Proposition, Propositions}, sibling=lem]{prop}
\declaretheorem[name=Definition, refname={Definition, Definitions}, Refname={Definition, Definitions}, style=definition, sibling=prop]{ddef}

\newcommand{\calO}{\mathcal{O}}

\DeclareMathOperator{\id}{Id}

\pgfplotsset{width=12cm, compat=1.9}

\hypersetup{
    pdftitle={IBM experiments},
    pdfauthor={Marcin Briański, Wojciech Burkot, Łukasz Czerwiński, Jan Gwinner, Vladyslav Hlembotskyi, Adam Szady}
}

\begin{document}
\title{Benchmarking 16-element quantum search algorithms\\on superconducting quantum processors}

\author{Jan~Gwinner}
\email{jan.gwinner@beit.tech}
\author{Marcin~Briański}
\email{marbri@beit.tech}
\author{Wojciech~Burkot}
\email{voytek@beit.tech}
\author{Łukasz~Czerwiński}
\email{lukasz@beit.tech}
\author{Vladyslav~Hlembotskyi}
\email{vlad@beit.tech}
\author{Adam~Szady}
\email{adsz@beit.tech}
\affiliation{Beit.tech}

\date{September 22, 2020}

\begin{abstract}
    We present experimental results on running 4-qubit unstructured search on IBM quantum processors. Our best attempt attained probability of success around 24.5\%. We try several algorithms  and use the most recent developments in quantum search to reduce the number of entangling gates that are currently considered the main source of errors in quantum computations. Comparing theoretical expectations of an algorithm performance with the actual data, we explore the hardware limits, showing sharp, phase-transition-like degradation of performance on quantum processors.  
    We conclude that it is extremely important to design hardware-aware algorithms and to include any other low level optimizations on NISQ devices.
\end{abstract}

    \maketitle
    
\section{Preliminaries}
\label{section:preliminaries}    
    
    In the \emph{unstructured search problem} we are given a phase oracle and want to find any marked element out of \(N\). The only action of the oracle is negating the amplitude of marked elements. This problem when considered on classical machines and classical oracles cannot be solved faster than in \(\Omega(N)\) oracle queries, but as showed by Grover in \cite{grover96} programmable quantum computers allow for an \(\calO(\sqrt{N})\) algorithm.
    
    There is an ongoing effort of implementing algorithms that solve the unstructured search problem
    on quantum computers. 
    We show how to solve this problem by utilizing small diffusion operators as is described 
    in \cite{groverfast}, \cite{wolfsearch}  and most recently in  \cite{prackum}. We present three successful implementations of unstructured search among \(16\) elements on IBM quantum computers.
    To the best of authors' knowledge, there has been no successful demonstration of quantum search in a space larger than \(8\) elements.
    
    Preparing efficient circuits for NISQ quantum computers requires acknowledgement of the topology of hardware. We have used hardware-aware circuits to improve previous results on IBM Q processors.

    \subsection{Prior work}\label{sub:prior_work}
        Since the invention of Grover's algorithm \cite{grover96}, there were plenty of attempts to run it on actual quantum hardware. So far, the largest search spaces on which the Grover's algorithm successfully and significantly amplified amplitude of the marked element were \(8\)-element spaces constructed on \(3\) qubits \cite{3q,kajman}. Some attempts to search for the marked element in a 16-element space were undertaken, see \cite{Mandviwalla,Stromberg}.  The results of \cite{Mandviwalla} are summarized in \Cref{tab:caption}. Back in 2018, es evidenced by the data therein,  quantum computers were unable to successfully run unstructured search in a space build on \(4\) qubits. Analysing these results one has to remember that the  probability of randomly finding one marked element among \(16\) in a classical setting is $6.25\%$.
        \begin{table}[htp] 
            \centering
            \begin{tabular}[t]{|l|r|r|r|} \toprule
                Algorithm (qubits used)& \# of gates & Accuracy & Execution time (s) \\ \midrule
                Grover 2-qubit (0,1)     & 18  & 74.05\% & 84.56 \\
                Grover 3-qubit (0,1,2)   & 33  & 59.69\% & 84.33 \\
                Grover 4-qubit (0,1,2,3) & 632 & 6.56\% &  185.13 \\\bottomrule
            \end{tabular}
            \caption{ \cite{Mandviwalla} results}
            \label{tab:caption}
        \end{table}
        We replicate some of previous results in order to understand the scale of hardware improvements achieved since the prior work has been completed.

    \subsection{Replication of prior work on current hardware}    
    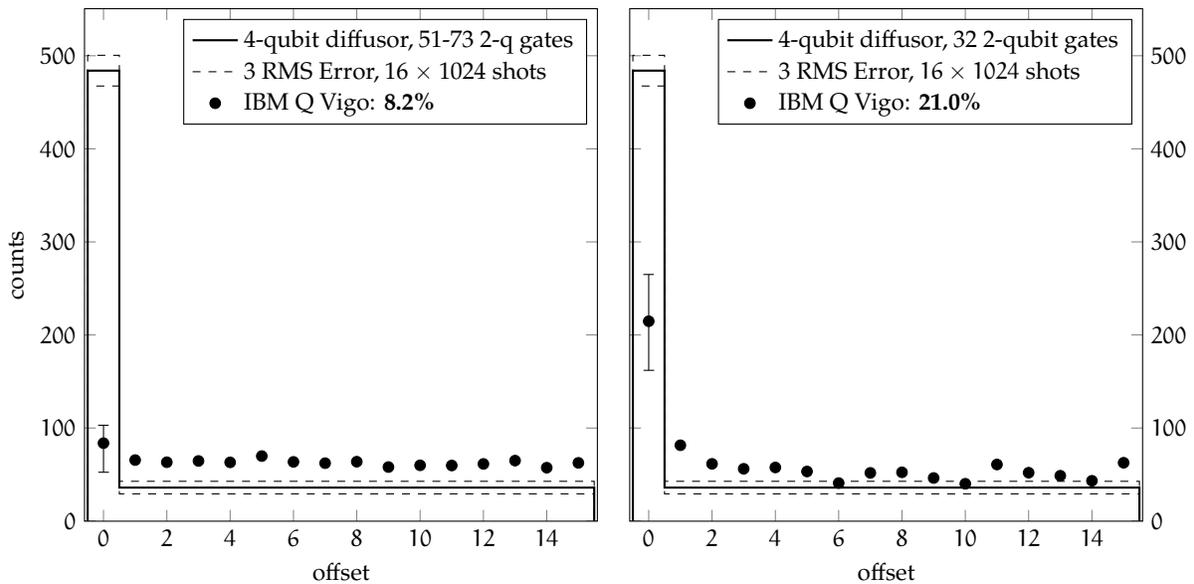
\begin{figure*}[htp]
        \begin{minipage}[b]{0.45\textwidth}
        \begin{tikzpicture}  
            \begin{axis}[
     	        height=8.4cm,
		        width=8.4cm,
	            xmin=-0.6,xmax=15.6,ymin=-0.1,
	            legend cell align=left,
	            xlabel={offset},
	            ylabel={counts}
	        ]
            \addplot [thick, mark=none] coordinates {
		        (-0.5,0) (-0.5,484)(0.5,484)
		        (0.5,36)(15.5,36)
		        (15.5,0)
	        };
            \addplot [dashed, mark=none] coordinates {	
 	            (-0.5,0)(-0.5,500.5)(0.5,500.5)
		        (0.5,42.75)(15.5,42.75)
		        (15.5,0)
            };
            \addplot [only marks,error bars, y dir=both, y explicit ] coordinates {
                (0,83.75) += (0,19.25) -=(0,31.25)
                (1,65.5625)
                (2,63.25)
                (3,64.625)
                (4,63.125)
                (5,69.875)
                (6,63.625)
                (7,62.25)
                (8,63.8125)
                (9,58.0625)
                (10,59.9375)
                (11,59.75)
                (12,61.4375)
                (13,65)
                (14,57.375)
                (15,62.5625)
            };
            \addplot [dashed, mark=none] coordinates {	
                (-0.5,0) (-0.5,467.5)(0.5,467.5)
		        (0.5,29.25)(15.5,29.25)
		        (15.5,0)
            };
            \addlegendentry{4-qubit diffusor, 51-73 2-q gates}
            \addlegendentry{3 RMS Error, 16 $\times$ 1024 shots}
            \addlegendentry{IBM Q Vigo: {\bf 8.2\%}}
            \end{axis}
        \end{tikzpicture}
        \end{minipage}
        \begin{minipage}[b]{0.45\textwidth}
        \begin{tikzpicture} 
	        \begin{axis}[
		        height=8.4cm,
    		    width=8.4cm,
	            xmin=-0.6,xmax=15.6,ymin=-0.1,
	            legend cell align=left,
	            xlabel={offset},
    	        ylabel near ticks, yticklabel pos=right
	        ]
            \addplot [thick, mark=none] coordinates {
		        (-0.5,0) (-0.5,484)(0.5,484)
    		    (0.5,36)(15.5,36)
	    	    (15.5,0)
	        };
            \addplot [dashed, mark=none] coordinates {	
 	            (-0.5,0)(-0.5,500.5)(0.5,500.5)
		        (0.5,42.75)(15.5,42.75)
    		    (15.5,0)
            };
            \addplot [only marks,error bars, y dir=both, y explicit ] coordinates {
                (0, 214.75) +=(0,50.25) -= (0, 52.75)
                (1, 81.5)
                (2, 61.5)
                (3, 56.1875)
                (4, 57.625)
                (5, 53.3125)
                (6, 40.875)
                (7, 51.8125)
                (8, 52.4375)
                (9, 46.3125)
                (10, 40.125)
                (11, 60.8125)
                (12, 52.0)
                (13, 48.75)
                (14, 43.3125)
                (15, 62.6875)
            };
            \addplot [dashed, mark=none] coordinates {	
     	        (-0.5,0) (-0.5,467.5)(0.5,467.5)
    		    (0.5,29.25)(15.5,29.25)
    		    (15.5,0)
            };
            \addlegendentry{4-qubit diffusor, 32 2-qubit gates}
            \addlegendentry {3 RMS Error, $16\times1024$ shots}
            \addlegendentry{IBM Q Vigo: {\bf 21.0\%}}
            \end{axis}
        \end{tikzpicture}
        \end{minipage}
        \caption{Comparison of results from running a single Grover's iteration on 4 qubits}
        \label{fig:grover 4qubit}
    \end{figure*}

        As the relaxation and dephasing times of real hardware have improved since 2018, we attempted to replicate the results from \cite{Stromberg} to investigate the improvements of hardware. 

        A straightforward reimplementation of \cite{Mandviwalla} of a single Grover's iteration on IBM Q Vigo yielded  the probability of finding  the marked element \(p_\text{succ}=8.2\%\) (averaged over all \(16\) oracles), while the lowest probability  was $5.1\%$, corresponding to the hardest oracle for the processor and the algorithm. The transpiller used  between $51$ and $ 73$ \(2\)-qubit gates, depending on the oracle. These results can be significantly improved. As \(2\)-qubit gate fidelities are noticably lower than their \(1\)-qubit counterparts, reducing the number of the former was our goal. Throughout this work, whenever we refer to 2-qubit gate count as a measure for the complexity of the circuit. 

\section{Our Results}
        Our implementation of a single iteration of Grover's algorithm used $32$ 2-qubit gates, counting native \(CNOT\) gates after transpilation. 
        The average  $p_\text{succ}$ was $21.0\%$. Moreover, our implementation had the desirable property of using the same number of \(CNOT\) gates independently of the oracle used.

        To avoid favouring any oracle in discussing the results, they are presented as the average  over possible bitwise symmetric differences between the measured element and the marked one, interpreted as numbers from \(0\) to  \(15\). Explicitly, for a given oracle marking the element \(\ket{x}\), where \(x \in \set{0, \dots, 15}\), whenever the measurement yielded \(\ket{y}\), where \(y \in \set{0, \dots, 15}\) we increment the count of \(y \oplus x \). This aligns the theoretical distributions, thus allowing us to aggregate the counts for different oracles. We are going to use the same approach when describing all our other results, unless clearly stated otherwise.
        \Cref{fig:grover 4qubit} presents the comparison between two implementations of a single iteration of Grover's algorithm. 
        \begin{table*}[htp]
            \begin{minipage}[b]{0.9\textwidth}
                \begin{tabular}[t]{|l|r|r|r|r|} \toprule
                    Algorithm (4 qubit search) & \# of 2-qubit gates & $p_\text{succ}$, average & $p_\text{succ}$, worst  \\ \midrule
                    Grover, 1 iteration, on IBM Q X5 \cite{Stromberg} \ & 51-73 & 6.62\% & (est.) 3\%  \\
                    same as above, on IBM Q Vigo   &51-73 & 8.2 \% & 5.1\% \\
                    same as above, optimized & 32& 21.0 \%& 15.8\%\\
                    \bottomrule
                \end{tabular}
            \end{minipage}
            \caption{ Replication of prior results on modern hardware with and without our improvements, $p_\text{succ}$ denotes probability of success averaged over all oracles}
            \label{tab:mandvilla_replication_results}
        \end{table*}  

        Each run of the algorithm consists of \(1024\) repetitions, which we sometimes call shots. We performed one run for each of  $16$ oracles.  
        The left graph in \Cref{fig:grover 4qubit} shows the results obtained using the implementation from~\cite{Stromberg}. The results of our optimized, topology-aware implementation are shown on the right side in the same figure. Both experiments were conducted on the IBM Q Vigo machine. This initial implementation forms a benchmark for future improvements. 
     
        \Cref{tab:mandvilla_replication_results} summarises the best results of running a single Grover's iteration on IBM Q Vigo, the last entry uses optimizations described in \cref{appendix:b}. Both the original and optimized implementations fail to attain the theoretical frequencies. In an absence of errors, $47.27\%$ of all measurements should yield the pattern corresponding to the oracle. More detailed discussion of the effects the decoherence has on the results is in \Cref{appendix:c}.
        
    \begin{figure*}[htp]  
        \begin{tikzpicture} 
	        \begin{axis}[
		        height=8.4cm,
		        width=8.4cm,
	            xmin=-0.6,xmax=15.6,ymin=-0.1,ymax=550,
	            legend cell align=left,
	            ylabel ={count}, 
	            xlabel near ticks, xticklabel pos=top
	        ]
            \addplot [thick, mark=none] coordinates {
                (-0.5,0)
		        (-0.5,400)
		        (0.5,400)
		        (0.5,16)
		        (1.5,16)
		        (3.5,16)
		        (3.5,144)
		        (4.5,144)
        		(4.5,16)
        		(7.5,16)
        		(7.5,144)
        		(8.5,144)
        		(8.5,16)
        		(11.5,16)
        		(11.5,144)
        		(12.5,144)
        		(12.5,16)
        		(15.5,16)
        		(15.5,0)
	        };
            \addplot [dashed, mark=none] coordinates {	
                (-0.5,0)
        		(-0.5,415)
        		(0.5,415)
        		(0.5,19)
        		(1.5,19)
        		(3.5,19)
        		(3.5,153)
        		(4.5,153)
        		(4.5,19)
        		(7.5,19)
        		(7.5,153)
        		(8.5,153)
        		(8.5,19)
        		(11.5,19)
        		(11.5,153)
        		(12.5,153)
        		(12.5,19)
        		(15.5,19)
        		(15.5,0)
        	};
            \addplot [only marks, error bars,y dir=both, y explicit] coordinates {
                (0, 186.625)     +=(0,78.375) -=(0,62.625)
                (1, 70.6875)
                (2, 50.75)
                (3, 36.6875)
                (4, 100.75)
                (5, 54.375)
                (6, 49.125)
                (7, 42.875)
                (8, 90.4375)
                (9, 39.1875)
                (10, 28.5625)
                (11, 28.6875)
                (12, 82.625)
                (13, 53.875)
                (14, 61.9375)
                (15, 46.8125)
            };
                (0, 395.6875)    +=(0,17.3125) -=(0, 21.6875)
                (1, 17.0625)
                (2, 14.6875)
                (3, 15.8125)
                (4, 146.75)
                (5, 15.375)
                (6, 17.4375)
                (7, 16.5)
                (8, 145.0)
                (9, 15.9375)
                (10, 17.0)
                (11, 14.625)
                (12, 145.1875)
                (13, 16.3125)
                (14, 14.5)
                (15, 16.125)
            \addplot [dashed, mark=none] coordinates {	
                (-0.5,0)
                (-0.5,385)
        		(0.5,385)
        		(0.5,13)
        		(1.5,13)
        		(3.5,13)
        		(3.5,135)
        		(4.5,135)
        		(4.5,13)
        		(7.5,13)
        		(7.5,135)
        		(8.5,135)
        		(8.5,13)
        		(11.5,13)
        		(11.5,135)
        		(12.5,135)
        		(12.5,13)
        		(15.5,13)
        		(15.5,0)
        	};
            \addlegendentry{2 2-qubit diffusors, 26 2-qubit gates}
            \addlegendentry {3 RMS Error, 16$\times$1024 shots}
            \addlegendentry{IBM Q Vigo {\bf 18.2\%}}
            \end{axis}
        \end{tikzpicture}
        \begin{tikzpicture} 
	        \begin{axis}[
		        height=8.4cm,
		        width=8.4cm,
	            xmin=-0.6,xmax=15.6,ymin=-0.1,ymax=550,
	            legend cell align=left,
	            ylabel near ticks, yticklabel pos=right,
	            xlabel near ticks, xticklabel pos=top
	        ]
            \addplot [thick, mark=none] coordinates {
                (-0.5,0)
        		(-0.5,256)
        		(0.5,256)
        		(0.5,64)
        		(1.5,64)
        		(3.5,64)
        		(3.5,0)
        		(4.5,0)
        		(4.5,64)
        		(7.5,64)
        		(7.5,0)
        		(8.5,0)
        		(8.5,64)
        		(11.5,64)
        		(11.5,0)
        		(12.5,0)
        		(12.5,64)
        		(15.5,64)
        		(15.5,0)
	        };
            \addplot [dashed, mark=none] coordinates {	
                 (-0.5,256) (-0.5,268) (0.5,268) (0.5,256)
                 (0.5,70) (3.5,70) (3.5,70) (3.5,0)(4.5,0)
                 (4.5,70)(4.5,70) (7.5,70) (7.5,64)(7.5,0)(8.5,0)
                 (8.5,70)(8.5,70) (11.5,70)(11.5,70) (11.5,0)(12.5,0)
                 (12.5,70)(12.5,70) (15.5,70) (15.5,70)(15.5,0)(-0.5,0)
                 (-0.5,244) (0.5,244)
                 (0.5,58) (3.5,58) (3.5,64) (3.5,0) (4.5,0)
                 (4.5,64)(4.5,58) (7.5,58) (7.5,64) (7.5,0)(8.5,0)
                 (8.5,64)(8.5,58) (11.5,58) (11.5,64)(11.5,0)(12.5,0)
                 (12.5,64)(12.5,58) (15.5,58) (15.5,64)
            };
            \addplot [only marks,black,  error bars, y dir=both, y explicit] coordinates {
                (0.2-0.3, 191.5625)    +=(0,28.4375) -=(0, 18.5625)
                (1.2-0.3, 62.9375)
                (2.2-0.3, 66.9375)
                (3.2-0.3, 68.8125)
                (4.2-0.3, 19.5)
                (5.2-0.3, 66.1875)
                (6.2-0.3, 70.375)
                (7.2-0.3, 57.375)
                (8.2-0.3, 16.375)
                (9.2-0.3, 56.375)
                (10.2-0.3, 67.0625)
                (11.2-0.3, 70.875)
                (12.2-0.3, 12.5)
                (13.2-0.3, 66.3125)
                (14.2-0.3, 66.9375)
                (15.2-0.3, 63.875)
            };
            \addplot [only marks,red,  error bars, y dir=both, y explicit] coordinates {
                (0, 173.75)      +=(0,22.25) -=(0,32.75)
                (1, 83.875)
                (2, 83.375)
                (3, 82.125)
                (4, 19.125)
                (5, 60.25)
                (6, 59.1875)
                (7, 63.25)
                (8, 20.125)
                (9, 49.375)
                (10, 47.0)
                (11, 53.5625)
                (12, 25.0)
                (13, 66.5)
                (14, 65.5625)
                (15, 71.9375)
            };
            \addplot [only marks,green, error bars, y dir=both, y explicit] coordinates {	
                (0.2, 138.625)     +=(0,27.375) -=(0,28.625)
                (1.2, 65.6875)
                (2.2, 72.8125)
                (3.2, 65.375)
                (4.2, 34.625)
                (5.2, 64.375)
                (6.2, 72.625)
                (7.2, 75.8125)
                (8.2, 33.1875)
                (9.2, 63.75)
                (10.2, 63.4375)
                (11.2, 71.0)
                (12.2, 19.6875)
                (13.2, 58.875)
                (14.2, 58.625)
                (15.2, 65.5)
            };
            \addlegendentry{2-qubit diffusor, 12 2-qubit gates}
            \addlegendentry{ 3 RMS Error,16 $\times$ 1024 shots}
            \addlegendentry{ {\bf 18.7\%} IBM Vigo, QV=16 }
            \addlegendentry{17.0\% IBM Q Valencia, QV=16}
            \addlegendentry{13.5\% IBM Q Essex, QV=8 }
            \addlegendentry{{\bf \hspace{0.5em}6.7\%} IBM Q Melbourne, QV=N/A}
            \end{axis}
        \end{tikzpicture}
        \begin{tikzpicture} 
	        \begin{axis}[
		        height=8.4cm,
        		width=8.4cm,
        	    xmin=-0.6,xmax=15.6,ymin=-0.1,ymax=550,
        	    legend cell align=left,
        	    ylabel={count},
        	    xlabel={offset}
	        ]
            \addplot [thick, mark=none] coordinates {
		        (-0.5,0) (-0.5,484)(0.5,484)
		        (0.5,36)(15.5,36)
		        (15.5,0)
	        };
            \addplot [dashed, mark=none] coordinates {	
 	            (-0.5,0)(-0.5,500.5)(0.5,500.5)
		        (0.5,42.75)(15.5,42.75)
		        (15.5,0)
            };
            \addplot [only marks,error bars, y dir=both, y explicit ] coordinates {
                (0, 214.75) +=(0,50.25) -= (0, 52.75)
                (1, 81.5)
                (2, 61.5)
                (3, 56.1875)
                (4, 57.625)
                (5, 53.3125)
                (6, 40.875)
                (7, 51.8125)
                (8, 52.4375)
                (9, 46.3125)
                (10, 40.125)
                (11, 60.8125)
                (12, 52.0)
                (13, 48.75)
                (14, 43.3125)
                (15, 62.6875)
            };
            \addplot [dashed, mark=none] coordinates {	
                (-0.5,0) (-0.5,467.5)(0.5,467.5)
		        (0.5,29.25)(15.5,29.25)
		        (15.5,0)
            };
            \addlegendentry{4-qubit diffusor, 28 2-qubit gates}
            \addlegendentry {3 RMS Error, $16\times1024$ shots}
            \addlegendentry{IBM Q Vigo: {\bf 21.0\%}}
            \end{axis}
        \end{tikzpicture}
        \begin{tikzpicture} 
	        \begin{axis}[
    		    height=8.4cm,
        		width=8.4cm,
    	        xmin=-0.6,xmax=15.6,ymin=-0.1,ymax=550,
    	        legend cell align=left,
        	    ylabel near ticks, yticklabel pos=right,
    	        xlabel={offset}
    	    ]
            \addplot [thick, mark=none] coordinates {
        		(-0.5,0)(-0.5,400)(0.5,400)
        		(0.5,64)(1.5,64)
        		(1.5,16)(2.5,16)
        		(2.5,64)(3.5,64)
        		(3.5,16)(4.5,16)
        		(4.5,64)(5.5,64)
        		(5.5,16)(6.5,16)
        		(6.5,64)(7.5,64)
        		(7.5,16)(8.5,16)
        		(8.5,66)(9.5,64)
        		(9.5,16)(10.5,16)
        		(10.5,64)(11.5,64)
        		(11.5,16)(12.5,16)
        		(12.5,64)(13.5,64)
        		(13.5,16)(14.5,16)
        		(14.5,64)(15.5,64)
        		(15.5,0)
        	};
            \addplot [dashed, mark=none] coordinates {	
    	        (-0.5,0)(-0.5,415)(0.5,415)
         		(0.5,70)(1.5,70)
        		(1.5,19)(2.5,19)
        		(2.5,70)(3.5,70)
        		(3.5,19)(4.5,19)
        		(4.5,70)(5.5,70)
        		(5.5,19)(6.5,19)
        		(6.5,70)(7.5,70)
        		(7.5,19)(8.5,19)
        		(8.5,70)(9.5,70)
        		(9.5,19)(10.5,19)
        		(10.5,70)(11.5,70)
        		(11.5,19)(12.5,19)
        		(12.5,70)(13.5,70)
        		(13.5,19)(14.5,19)
        		(14.5,70)(15.5,70)
        		(15.5,0)
        	};
            \addplot [only marks,error bars, y dir=both, y explicit ] coordinates {
                (0, 251.375)     +=(0,46.625) -=(0, 40.375)
                (1, 75.875)    
                (2, 43.5)
                (3, 50.4375)
                (4, 41.9375)
                (5, 70.0625)
                (6, 32.0)
                (7, 44.8125)
                (8, 61.0)
                (9, 47.625)
                (10, 25.375)
                (11, 56.9375)
                (12, 37.25)
                (13, 58.625)
                (14, 48.25)
                (15, 78.9375)
            };
            \addplot [only marks, green,error bars, y dir=both, y explicit ] coordinates {
                (0.2, 187.125)  += (0,20.875) -=(0, 38.125)
                (2.2, 50.3125)
                (1.2, 77.5)
                (3.2, 71.25)
                (4.2, 37.4375)
                (6.2, 33.625)
                (5.2, 69.0625)
                (7.2, 68.125)
                (8.2, 53.4375)
                (10.2, 26.3125)
                (9.2, 56.1875)
                (11.2, 69.75)
                (12.2, 31.875)
                (14.2, 34.4375)
                (13.2, 72.0)
                (15.2, 85.5625)
            };
            \addplot [dashed, mark=none] coordinates {	
 	            (-0.5,0) (-0.5,385)(0.5,385)
        		(0.5,58)(1.5,58)
        		(1.5,13)(2.5,13)
        		(2.5,58)(3.5,58)
        		(3.5,13)(4.5,13)
        		(4.5,58)(5.5,58)
        		(5.5,13)(6.5,13)
        		(6.5,58)(7.5,58)
        		(7.5,13)(8.5,13)
        		(8.5,58)(9.5,58)
        		(9.5,13)(10.5,13)
        		(10.5,58)(11.5,58)
        		(11.5,13)(12.5,13)
        		(12.5,58)(13.5,58)
        		(13.5,13)(14.5,13)
        		(14.5,58)(15.5,58)
        		(15.5,0)
            };
            \addlegendentry{3-qubit diffusor, 24 2-qubit gates}
            \addlegendentry {3 RMS Error, $16 \times 1024$ shots}
            \addlegendentry{IBM Q Vigo, {\bf 24.5\%}}
            \addlegendentry{IBM Q Vigo, 18.2\%}
            \end{axis}
        \end{tikzpicture}
        \caption{Theoretical and experimental results on runs of four algorithms, each finding a single marked element in a space of \(16\) elements, differing in the size of the diffusion operators and the number of iterations}
        \label{Fig:various}
    \end{figure*}
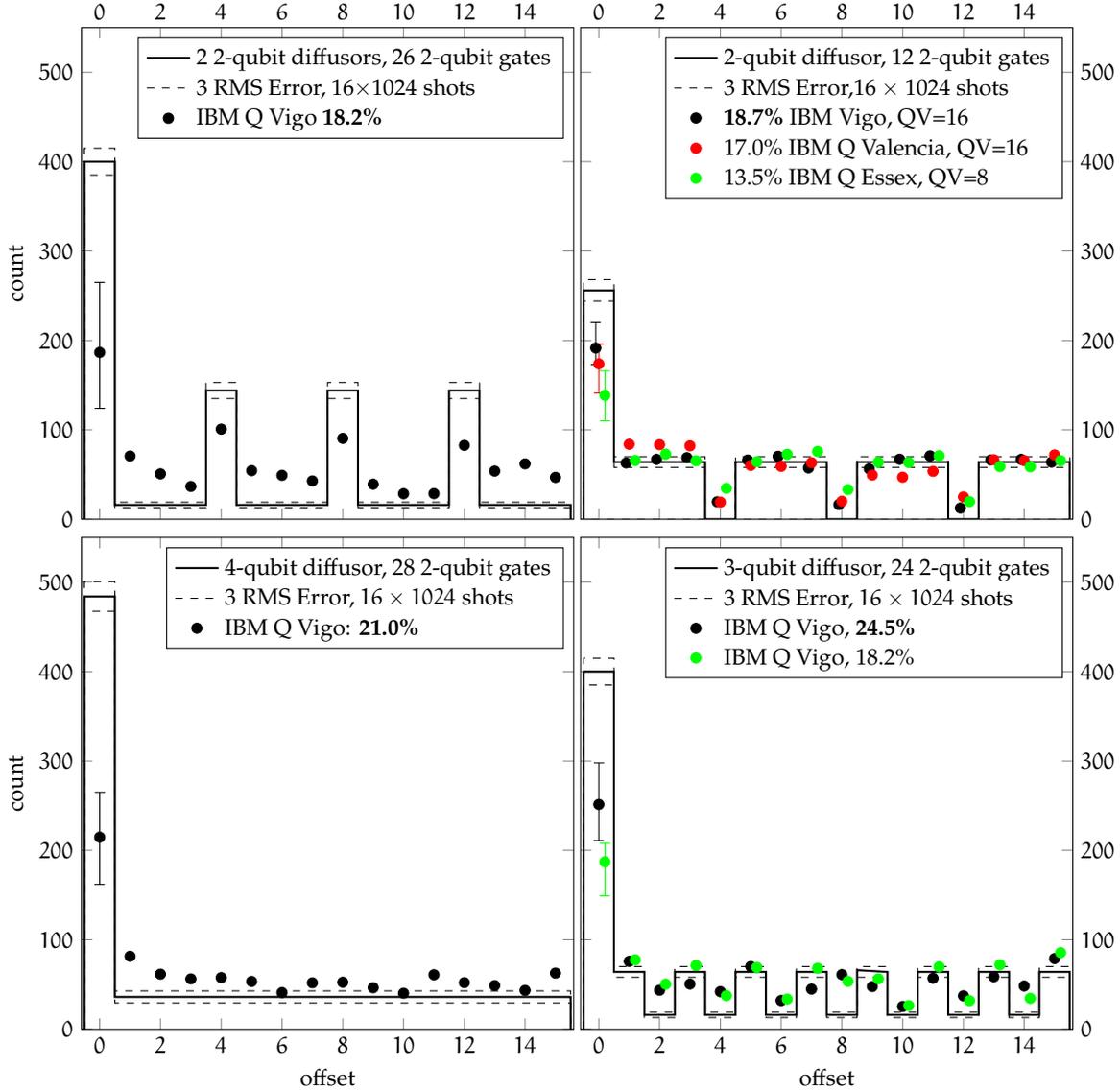 

    Our aim was to find algorithms and implementation methods for NISQ processors, suitable for demonstration of abilities of these machines to search for a single element in \(16\)-element space.  
    As the candidates, we implement the first iteration of unstructured search algorithms,
    employing \(4\)-qubit oracles for $16$ possible search patterns and \(2\)-, \(3\)- and \(4\)-qubit diffusion operators. Additionally, using an approach similar to \cite{prackum}, we perform a fragment of optimal quantum unstructured search algorithm (although using a slightly different pattern of diffusion operators, see \Cref{def:Drzewker}). Implementing oracles and diffusion operators we take care to implement them with accordance to topology of quantum hardware, for more details see \Cref{appendix:b}. To further reduce the number of \(2\)-qubit gates we use the technique of partial uncompute that allows us to not uncompute some ancillae but keep them to expedite the next oracle call; it is described in detail in \cite{prackum}. 
        
    \subsection{The main result}
        The main result of this paper, to the best of authors' knowledge, is \emph{the first} demonstration of quantum unstructured search in \(16\)-element space, yielding statistically significant outcome on actual quantum computing hardware. Four different algorithms, differing in the size of the diffusion operators and the number of iterations were run on processors from IBM Q family. The results are summarised in  \Cref{Fig:various}. The upper plots show prefixes of $D_2$ circuits for $\overline{k} = (2, 2)$. The lower left plot shows a single iteration of Grover's algorithm, and the lower right plot presents a partial search with 3-qubit diffusor. The plots are ordered by their measured probability of success $p_\text{succ}$. The range of $p_\text{succ}$ is from $18.2\%$ to $24.5\%$. The numbers of 2-qubit gates of the implementations vary from $12$  for a single iteration of search using \(2\)-qubit diffusion operator to $28$ for a single iteration of Grover's algorithm. The only implementation of the algorithm with multiple oracle queries presented shows relatively low $p_\text{succ}=18.2\%$, as it introduces a new source of errors, absent in variants with a single oracle query. Even the best result fails (albeit by a narrow margin) to attain the expected number of oracle queries better that the classical random search.

        Results for a selection of processors running a single iteration of unstructured search using \(2\)-qubit diffusion operator are presented in \Cref{Fig:various} top right plot. We have selected IBM Q Vigo to perform longer circuits. The results of these runs are summarised in the other plots of \Cref{Fig:various}.

    \subsection{Efficiency of NISQ hardware}

\begin{figure*}[htp] 
            \begin{tikzpicture}
	            \begin{axis}[
		            height=10cm,
		            width=10cm,
	                xmin=0,xmax=70, ymin=0, ymax=1,
	                legend pos=north east, 
	                legend cell align = left,
	                xlabel ={number of 2-qubit gates},
	                ylabel ={$R$}
	            ]
                \addplot [only marks, error bars, y dir=both, y explicit] coordinates {
                    (14,157*1.024/256) +- (0.0,0.0501)
                    (28,208*1.024/400) +- (0.0,0.037)
                    (30,162*1.024/400) +- (0,0.032)
                    (32,124*1.024/484) +- (0,0.024)
                    (62,82*1.024/484) +- (0.0,0.019)
                    (28,.444) +- (0,0.0335)
                    (12,.748) +- (0,0.06)   
                };
                \addplot [only marks,mark=o,error bars, y dir=both, y explicit] coordinates {
                    (26,.465) +- (0,0.0345) 
                    (38,.1895) +- (0,0.016) 
                };
                \addplot [only marks,green,error bars, y dir=both, y explicit] coordinates { 
                    (24,.627) +- (0,0.05)
                };
                \addplot[dashed, domain = 0:70] {.0625+.9375*.94*(exp(x*ln(0.97)))};
                \addplot[thick,red,domain = 0:70] {0.16969+0.498/(1 + exp(x*(0.3617)-10.459))};
                \addlegendentry{search with 1 oracle query}
                \addlegendentry{ search with $>1$ oracle queries}
                \addlegendentry{best $p_\text{succ}$, \(3\)-qubit diffusor}
                \addlegendentry{ best fit for fixed gate fidelities}
                \addlegendentry{ best fit logistic curve }
                \end{axis}
            \end{tikzpicture}
            \caption{IBM Q Vigo efficiency}{The ratio $R$ of actual to theoretical success frequencies for different algorithms selecting 1 of 16 states vs. 2-qubit gate counts of these algorithms for IBM Q Vigo (errors are \(1\) RMSE)}
            \label{fig:efficiency}
        \end{figure*}
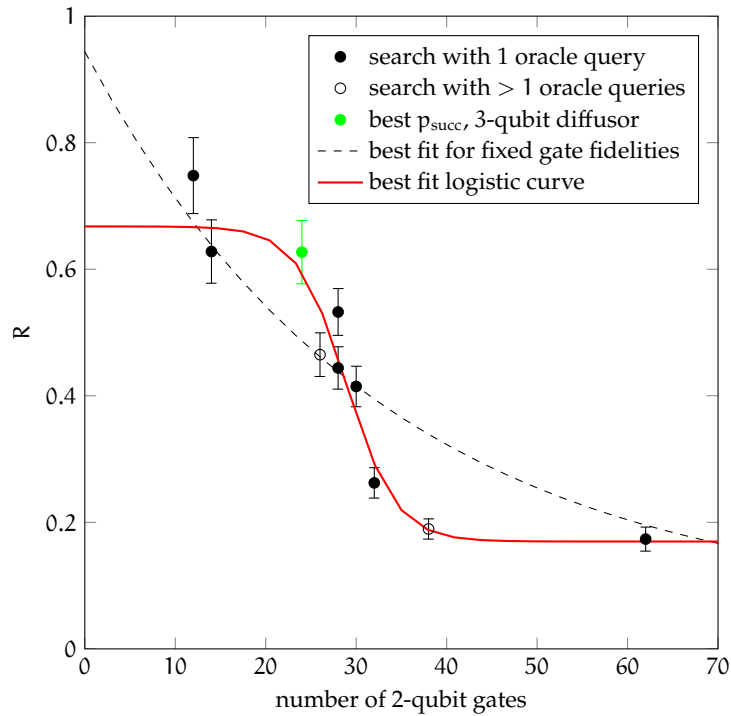

        Let us define $R$ as a ratio of actually achieved frequency of counts and the theoretical probability of success $p_\text{succ}$ of an algorithm, searching for a single marked element out of 16. Such defined $R$ is presented in  \Cref{fig:efficiency}, pointing to a sharp degradation of the performance of IBM Q Vigo at \(2\)-qubit gate count of about 30. 

        The dotted line in \cref{fig:efficiency} denotes the best fit estimation of degrading performance caused by infidelities of 2-qubit quantum gates as well as setup and measurement errors. It seems not to be enough to explain the behaviour of our algorithms, as the efficiency of those circuits when compared to theoretical results conforms to the red line which is best fit logistic curve.
        We show that current quantum hardware favours short circuits, as two steps of \(D_2\) algorithm that should yield the same probability of measuring the marked element as \(3\)-qubit partial search and higher probability than a single application of \(2\)-qubit diffusor yielded worse results.  
        Besides efficient implementations, many of these results would not be possible if not for consideration of smaller diffusion operators. These were first introduced by Grover in \cite{groverfast} and later explored in \cite{wolfsearch} and \cite{prackum}. The concept of benefits arising from the use of local diffusion operators has been studied in other papers, e.g. \cite{zhang}.
        
        Additionally, we demonstrate full search space entanglement. While a single application of a \(2\)-qubit diffusor entangles just \(\sfrac{1}{4}\) of states, \(3\)-qubit diffusor entangles half of the states, applying \(4\)-qubit diffusor or two \(2\)-qubit diffusors entangles all the states in the search space as seen in \Cref{Fig:various} 

    \subsection{Further remarks}
        This paper shows that it is extremely important to design quantum algorithms on modern NISQ devices with the awareness of their topology  to achieve the best performance possible. Further development can be aimed in exploration of better oracle implementation or replacing known algorithms for solving the unstructured search problem with ones more suitable for a given hardware architecture. It can be noted that the placement of diffusion operators in our best circuits is not accidental but carefully chosen among all other possibilities. This is of course possible only due to the fact that we may, most of the time, forgo diffusion operators that act on all qubits. 
    
\section{Methodology and Implementation}
    
    Firstly, we present $D_n$ circuit
    that is constructed in similar manner to \(W_n\) from \cite{prackum}, but forces higher amplitude of the marked element then circuit \(W_n\) during the first three steps. Circuit \(D_n\) also allows for the construction of optimal circuits as stated in Appendix A in \cite{prackum}. In the following definition we adapt the notation from the aforementioned work.
    \begin{ddef}\label{def:Drzewker}
        Let $\overline{k} = (k_1, \dots, k_m)$ be a sequence of positive integers and let $n := \sum_{j=1}^m k_j$. Given a quantum oracle $O$, for $j \in \set{0, \dots, m}$ we define the circuit $D_j$ recursively as follows:
        \begin{align*}
            D_0 &= \id_{n}\\
            D_{j + 1} &= D_{j}\paren*{\id_{k_1 + \dots + k_{j}} \otimes G_{k_{j + 1}} \otimes \id_{k_{j + 2} + \dots + k_m}}OD_{j} \text{.}
        \end{align*}
    \end{ddef}
    
    \noindent
    In \cite{prackum} we prove that \(D_n\) circuits with Amplitude Amplification \cite{ampamp} indeed allow us to perform  optimal quantum search. The $D_n$ circuits have multiple benefits over Grover's algorithm. They use smaller diffusion operators which require fewer number of elementary gates to implement. The $D_n$ circuits are quite flexible and can be implemented in a topology-aware and hardware-aware manner. The sparse structure of \(D_n\) allows some of the ancillae used in implementation of oracles to stay not uncomputed, as explained in \cite{prackum} in section devoted to the partial uncompute technique.
    
    Secondly, we also try to implement the oracle as efficiently as possible. Notice that if we have an ancilla qubit we can decompose standard $CCCX$ gate into two $CCX$ and one $CCZ$ gates. The second $CCX$ is basically needed solely to uncompute the byproduct on the ancilla qubit. We notice that it is possible to apply partical uncompute technique \cite{prackum}, so sometimes we can spare the second $CCX$. Additionally, we can replace the first $CCX$ with Margolus gate \cite{margolus}. The implementation can be seen in \Cref{fig:moracle}, the implementation of the first step of Grover's algorithm can be seen in \Cref{fig:topology}.

    Thirdly, it is crucial to be topology-aware and hardware-aware when implementing the circuits. This way it is possible to achieve drastic improvement of performance on NISQ devices. As it was mentioned in \Cref{section:preliminaries}, it is possible to achieve major improvement on IBM Q Vigo by being hardware-aware.
    
    \begin{acknowledgements}
        We would like to express our deep gratitude to our friends at Beit, in particular to Jacek Kurek, for their insights and criticism.
        However, mere language would not suffice for this endeavour, so we will refrain from doing so.
    \end{acknowledgements}
 
    \clearpage   
    \makeatletter\onecolumngrid@push\makeatother
    
    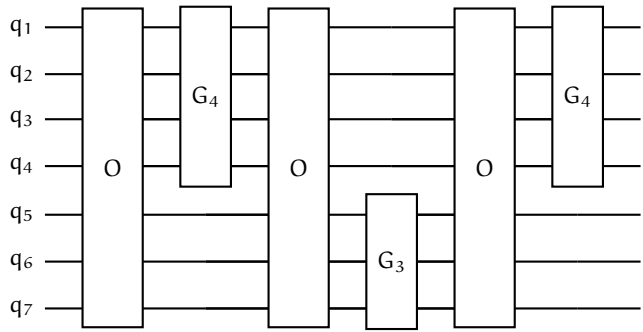
\begin{figure*}[htp]
        \begin{minipage}[b]{0.9\textwidth}
            \begin{quantikz}[row sep=0.1cm, column sep=0.5cm]
                \lstick{$q_1$} & \gate[7][0.8cm]{O} & \gate[4]{G_4} &  \gate[7][0.8cm]{O} & \qw           & \gate[7][0.8cm]{O} & \gate[4]{G_4} & \qw \\
                \lstick{$q_2$} & \ghost{O}          & \qw           & \qw                & \qw           & \qw                & \qw           & \qw \\
                \lstick{$q_3$} & \ghost{O}          & \qw           & \qw                & \qw           & \qw                & \qw           & \qw \\ 
                \lstick{$q_4$} & \ghost{O}          & \qw           & \qw                & \qw           & \qw                & \qw           & \qw \\
                \lstick{$q_5$} & \ghost{O}          & \qw           & \qw                & \gate[3]{G_3} & \qw                & \qw           & \qw \\ 
                \lstick{$q_6$} & \ghost{O}          & \qw           & \qw                & \qw           & \qw                & \qw           & \qw \\
                \lstick{$q_7$} & \ghost{O}          & \qw           & \qw                & \qw           & \qw                & \qw           & \qw \\
            \end{quantikz}
        \end{minipage}
        \caption{$D_2$ for $\overline{k} = (4, 3)$}
        \label{fig:drzewker}
    \end{figure*}
    
    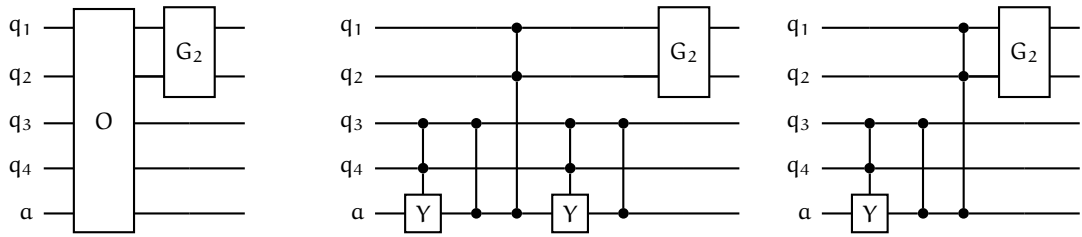
\begin{figure*}
        \begin{minipage}[b]{0.3\textwidth}
            \begin{quantikz}[row sep=0.1cm, column sep=0.4cm]
                \lstick{$q_1$} & \gate[5][0.8cm]{O} & \gate[2]{G_2} & \qw \\
                \lstick{$q_2$} & \ghost{O}          & \qw           & \qw \\
                \lstick{$q_3$} & \ghost{O}          & \qw           & \qw \\ 
                \lstick{$q_4$} & \ghost{O}          & \qw           & \qw \\
                \lstick{$a$}   & \ghost{O}          & \qw           & \qw \\
            \end{quantikz}
        \end{minipage}
        \begin{minipage}[b]{0.3\textwidth}
            \begin{quantikz}[row sep=0.1cm, column sep=0.4cm]
                \lstick{$q_1$} & \qw            & \qw           & \ctrl{1}          & \qw & \qw                 & \gate[2]{G_2} & \qw & \ghost{O} \\
                \lstick{$q_2$} & \qw            & \qw           & \ctrl{3}          & \qw & \qw & \qw                 & \qw         & \ghost{O} \\
                \lstick{$q_3$} & \ctrl{1}       & \ctrl{2}      & \qw               & \ctrl{1} & \ctrl{2} & \qw       & \qw          & \ghost{O} \\
                \lstick{$q_4$} & \ctrl{1}       & \qw           & \qw               & \ctrl{1} & \qw & \qw            & \qw     & \ghost{O} \\
                \lstick{$a$}   & \gate{Y}       & \control{}    & \control{}        & \gate{Y} & \control{} & \qw     & \qw            & \ghost{O} \\
            \end{quantikz}
        \end{minipage}
        \begin{minipage}[b]{0.3\textwidth}
            \begin{quantikz}[row sep=0.1cm, column sep=0.4cm]
                \lstick{$q_1$} & \qw            & \qw           & \ctrl{1}          & \gate[2]{G_2}  & \qw                 & \ghost{O} \\
                \lstick{$q_2$} & \qw            & \qw           & \ctrl{3}          & \qw            & \qw                 & \ghost{O} \\
                \lstick{$q_3$} & \ctrl{1}       & \ctrl{2}      & \qw               & \qw            & \qw                 & \ghost{O} \\
                \lstick{$q_4$} & \ctrl{1}       & \qw           & \qw               & \qw            & \qw                 & \ghost{O} \\
                \lstick{$a$}   & \gate{Y}       & \control{}    & \control{}        & \qw            & \qw                 & \ghost{O} \\
            \end{quantikz}
        \end{minipage}
        \caption{The first step of $D_2$ for $\overline{k} = (2, 2)$ with one ancilla qubit.}
        
        \label{fig:moracle}
    \end{figure*}
    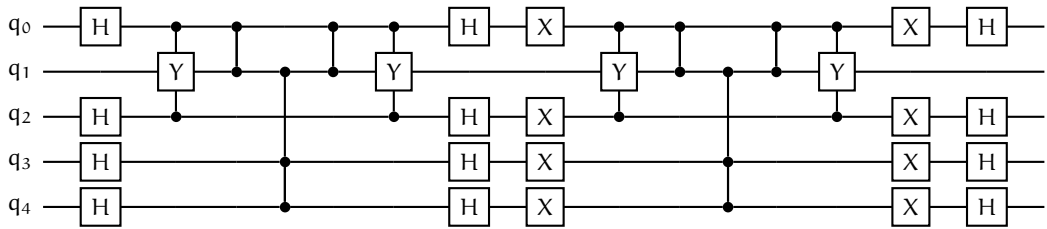
\begin{figure*}[htp]
        \begin{quantikz}[row sep=0.1cm, column sep=0.5cm]
            \lstick{$q_0$}& \gate{H} & \ctrl{1} & \ctrl{1}& \qw & \ctrl{1}     & \ctrl{1} & \gate{H}&\gate{X}& \ctrl{1} & \ctrl{1}& \qw & \ctrl{1}     & \ctrl{1} & \gate{X}&\gate{H}&\qw \\
            \lstick{$q_1$}& \qw & \gate{Y} & \ctrl{-1}& \ctrl{2} &  \ctrl{-1}  & \gate{Y}&\qw&  \qw& \gate{Y} & \ctrl{-1}& \ctrl{2} &  \ctrl{-1}  & \gate{Y}&\qw&  \qw &\qw         \\
            \lstick{$q_2$}&\gate{H} & \ctrl{-1} & \qw& \qw & \qw      &  \ctrl{-1}&\gate{H} &  \gate{X}& \ctrl{-1} & \qw& \qw & \qw      &  \ctrl{-1}&\gate{X} &  \gate{H}  &\qw      \\ 
            \lstick{$q_3$}& \gate{H} & \qw    & \qw &  \ctrl{1}  &  \qw & \qw &\gate{H}& \gate{X}& \qw    & \qw &  \ctrl{1}  &  \qw & \qw &\gate{X}& \gate{H}      &\qw     \\
            \lstick{$q_4$}& \gate{H} & \qw    & \qw &  \ctrl{-1} & \qw & \qw &\gate{H} & \gate{X}& \qw    & \qw &  \ctrl{-1} & \qw & \qw &\gate{X} & \gate{H}      &\qw  \\
        \end{quantikz}
        \caption{ Implementation of the first step of Grover's algorithm on qubits (0,2,3,4) on IBM Q Vigo quantum processor}
        \label{fig:topology}
    \end{figure*}
    \clearpage
    \makeatletter\onecolumngrid@pop\makeatother

    \addcontentsline{toc}{section}{References}
    \bibliography{references}

    \include{TopologyAppendix}

    \include{kullback}
\end{document}

%% file: TopologyAppendix.tex
\section{Topology-Aware Implementation}\label{appendix:b}

\begin{figure}[htp]
        \centering
        \includegraphics[scale=0.4]{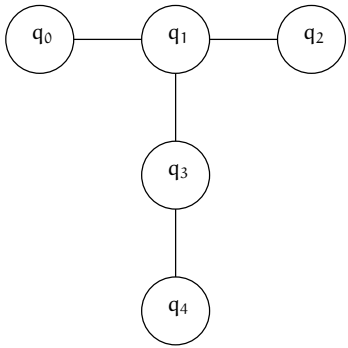}
        \caption{Topology of IBM Q Vigo quantum processor; edges denote pairs of qubits on which 2-qubit gates can be applied}
        \label{fig:figa_zmakiem}
    \end{figure}

    Most of decoherence in hardware comes from \(CNOT\) gates, so the hardware-aware optimizations performed by us were focused mostly on reducing their number.
    IBM software transpiles any quantum circuit to an equivalent one that consists of arbitrary 1-qubit and $CNOT$ gates. Moreover, there are  restrictions on which pairs of qubits a $CNOT$ gate can be applied to, in this case \(SWAP\) gates are used to transport the relevant qubits to the suitable pair of qubits adjacent in the underlying topology. Similar methods were utilised in~\cite{htris}. Each of the \(SWAP\) gates requires \(3\) \(CNOT\) gates to be implemented. These restrictions vary from one quantum computer to another. For example, IBM Q Vigo has these restrictions as in \Cref{fig:figa_zmakiem}. By careful analysis of architecture of IBM quantum computers we reduce the number of $CNOT$ gates in our circuits noticeably. Let us note that restrictions from \cref{fig:figa_zmakiem} apply also to Valencia, Ourense and Essex quantum processors.

    Let us restrict ourselves to the IBM Q Vigo topology and try to implement a quantum circuit from \cite{prackum} as close to optimal as possible. First, let us try to optimize the total number of $CNOT$ gates in oracle (see \Cref{fig:moracle}). We can use qubit \(q_1\) as a target so that all \(2\)-qubit gates needed to perform Margolus gate can be run on the neighbouring qubits, see \cref{fig:margolus}.
    Margolus gate is a substitute for a standard Toffoli gate whenever we only aggregate result of logical \(AND\) operation between two input qubits (in our case \(q_0, q_2\)) in ancilla qubit (i.e. \(q_1\)). Furthermore, it is usually required to uncompute this operation before proceeding with further computations that involve input qubits.

\begin{figure*}[htp]
        \begin{minipage}[b]{0.9\textwidth}
            \begin{quantikz}[row sep=0.1cm, column sep=0.4cm]
        	 	\lstick{ ${q}_{1}$  } & \ctrl{1} & \qw & \qw & \qw & \ctrl{1} & \gate[1][1.8em]{T} & \qw & \qw & \swap{} & \qw & \qw & \qw & \swap{} & \qw & \qw\\
        	 	\lstick{ ${q}_{3}$  } & \targ{} & \gate[1][1.8em]{T^\dag} & \targ{} & \gate[1][1.8em]{T} & \targ{} & \gate[1][1.8em]{T^\dag} & \targ{} & \gate[1][1.8em]{T} & \swap{-1} & \targ{} & \gate[1][1.8em]{T^\dag} & \targ{} & \swap{-1} & \qw & \qw\\
        	 	\lstick{ ${q}_{4}$  } & \qw & \qw & \ctrl{-1} & \qw & \qw & \qw & \ctrl{-1} & \qw & \qw & \ctrl{-1} & \gate[1][1.8em]{T} & \ctrl{-1} & \qw & \qw & \qw\\
        	 \end{quantikz}
    	 \end{minipage}
        \caption{Na\"ive implementation of \(CCZ\) on qubits \(q_1, q_3, q_4\) in line topology}
        \label{fig:swap}
    \end{figure*}
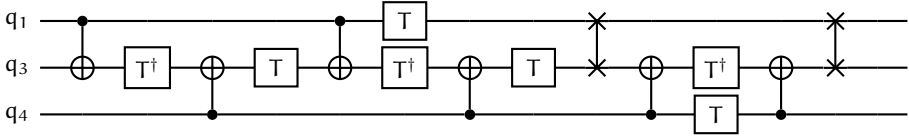

    To implement a \(CCZ\) gate on qubits \(q_1, q_3, q_4\) with standard approach we would need to perform two \(SWAP\)s to let qubits \(q_1\) and \(q_4\) interact, see \Cref{fig:swap}
    
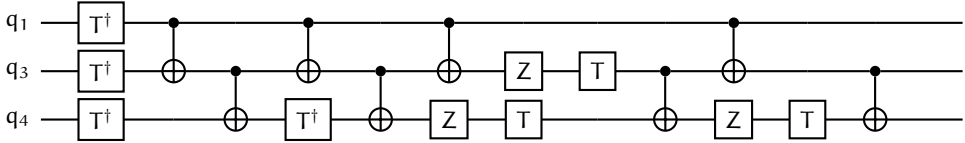
\begin{figure*}[htp]
        \begin{minipage}[b]{0.9\textwidth}
            \begin{quantikz}[row sep=0.1cm, column sep=0.5cm]
        	 	\lstick{ ${q}_{1}$ } & \gate{T^\dag} & \ctrl{1} & \qw & \ctrl{1} & \qw & \ctrl{1} & \qw & \qw & \qw & \ctrl{1} & \qw & \qw & \qw & \qw\\
        	 	\lstick{ ${q}_{3}$ } & \gate{T^\dag} & \targ{} & \ctrl{1} & \targ{} & \ctrl{1} & \targ{} & \gate{Z} & \gate{T} & \ctrl{1} & \targ{} & \qw & \ctrl{1} & \qw & \qw\\
        	 	\lstick{ ${q}_{4}$ } & \gate{T^\dag} & \qw & \targ{} & \gate{T^\dag} & \targ{} & \gate{Z} & \gate{T} & \qw & \targ{} & \gate{Z} & \gate{T} & \targ{} & \qw & \qw\\
        	 \end{quantikz}
    	 \end{minipage}
        \caption{Simplified implementation of \(CCZ\) on qubits \(q_1, q_3, q_4\) in line topology}
        \label{fig:swap_simplified_better}
    \end{figure*}

    It can be circumvented with a more efficient circuit that requires only \(8\) \(CNOTs\), see \Cref{fig:swap_simplified_better}.
    
\begin{figure*}[htp]
        \begin{minipage}[b]{0.45\textwidth}
        \subfloat[Na\"ive implementation -- 7 \(CNOT\)s]{%
             \begin{quantikz}[row sep=0.1cm, column sep=0.4cm]
    	 	\lstick{ ${q}_{0}$  } & \gate{H} & \gate{X} & \swap{1} & \qw & \swap{1} & \gate{X} & \gate{H} & \qw & \qw\\
         	\lstick{  $\ket{0}$  } & \qw & \qw & \swap{-1} & \ctrl{1} & \swap{-1} & \qw & \qw & \qw & \qw\\
         	\lstick{ ${q}_{2}$  } & \gate{H} & \gate{X} & \qw & \control{}\qw & \qw & \gate{X} & \gate{H} & \qw & \qw\\
    	 \end{quantikz}}
    	 \end{minipage}
        \begin{minipage}[b]{0.45\textwidth}
        \subfloat[Improved implementation -- 3 \(CNOT\)s]{%
            \begin{quantikz}[row sep=0.1cm, column sep=0.4cm]
    	 	\lstick{ ${q}_{0}$  } & \gate{H} & \gate{X} & \ctrl{1} & \qw & \ctrl{1} & \gate{X} & \gate{H} & \qw & \qw\\
         	\lstick{ $\ket{0}$  } & \qw & \qw & \targ{} & \ctrl{1} & \targ{} & \qw & \qw & \qw & \qw\\
         	\lstick{ ${q}_{2}$  } & \gate{H} & \gate{X} & \qw & \control{}\qw & \qw & \gate{X} & \gate{H} & \qw & \qw\\
    	 \end{quantikz}}
    	 \end{minipage}
        \caption{Diffusor on qubits \(q_0,q_2\) that are topologically separated by qubit \(q_1\) that is in state \( \ket{0}\)}
        \label{fig:better_diffusor}
    \end{figure*}
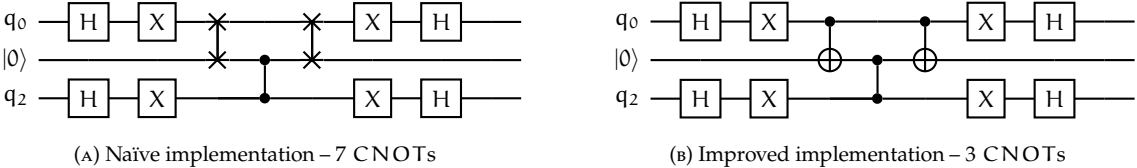

    We also reduce the total number of $CNOT$ gates in implementations of diffusors. Implementation of a diffusor on qubits \(q_3, q_4\) is straightforward and costs only \(1\) entangling gate. To implement a diffusor on qubits \(q_0, q_2\) we would need to use two \(SWAP\) gates that would result in a circuit with \(7\) entangling gates. To reduce the amount of entangling gates we note that there exists an equivalent circuit that requires only three 2-qubit gates, see \Cref{fig:better_diffusor}.
    
    \clearpage

    \makeatletter\onecolumngrid@push\makeatother
    
    \begin{figure*}[htp]
        \begin{minipage}[b]{0.9\textwidth}
            \begin{quantikz}[row sep=0.1cm, column sep=0.5cm]
        	 	\lstick{ ${q}_{0}$ } & \qw & \qw & \qw & \ctrl{1} & \qw & \qw & \qw & \qw & \qw & \ghost{O}\\
        		 	\lstick{ ${q}_{1}$ } & \gate{R_1} & \targ{} & \gate{R_1} & \targ{} & \gate{R_2} & \targ{} & \gate{R_2} & \qw & \qw & \ghost{O}\\
        	 	\lstick{ ${q}_{2}$ } & \qw & \ctrl{-1} & \qw & \qw & \qw & \ctrl{-1} & \qw & \qw & \qw & \ghost{O}\\
        	 \end{quantikz}
    	 \end{minipage}
        \caption{Margolus gate, where $R_1 = U_3(\frac{\pi}{4},0,0)$ and $R_2 = U_3(-\frac{\pi}{4},0,0)$}
        \label{fig:margolus}
    \end{figure*}
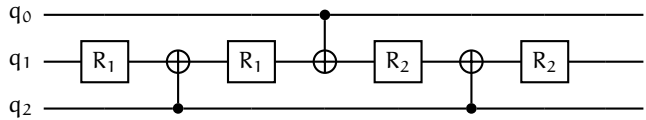
    
    \clearpage
    \makeatletter\onecolumngrid@pop\makeatother
    
    
    

%% file: kullback.tex
\section{Error Mitigation}\label{appendix:c}

The performance of different quantum processors in IBM Q family varies significantly. In \Cref{fig:mel vs vigo} we compare IBM Q Melbourne against IBM Q Vigo, running a single iteration of unstructured search in \(16\)-element space using \(2\)-qubit diffusion operators, implemented with a 2-qubit gate count of $16$, as an example.  We run 1024 shots of this algorithm for each of $16$ possible oracles.

While IBM Q Melbourne failed to find the marked element, IBM Q Vigo attained \(p_\text{succ} =15\%\).  We investigated the difference in performance with the following approach.

In the main body of this work, we used the method described in \Cref{sub:prior_work}, which allowed for superimposing the distributions for different oracles. 
Contrary, in this Appendix, we add the counts for each measured pattern, effectively measuring  the quantum processor's preference for each measured pattern, while not affecting the 2-qubit circuit depths of experiments.

The biggest effect is related to decoherence, bringing qubits to their ground state. This results in patterns containing more \(0\)'s to be observed more often, as presented in \Cref{fig:mel vs vigo}, where the colours code the distance from the expected average value of counts, marked in dashed line.

The same effect becomes evident for IBM Q Vigo for larger circuits, for example for a single iteration of the \(3\)-qubit diffusor  as illustrated in \Cref{fig:Vigo decoherence}, the left hand side plot. The right side plot shows dependence on the state of Lowest Significant Qubit of the pattern. We have attempted mitigating these errors, by computing a $16\times16$ correction matrix, using counts we measured for all the patterns for all the oracles. Subsequently, we have applied the corrections to raw counts, reshuffled as for  \Cref{Fig:various}, so the theoretical distributions overlap. After the correction, the average $p_\text{succ}$ changed marginally from $24.5\%$ to $ 24.8\%$ (still not enough to better a classical search in an expected number of oracle calls), however the result for the worst performing oracle improved from $20.6\%$ to $22.7\%$.

\clearpage

\makeatletter\onecolumngrid@push\makeatother
    \begin{figure*}[htp]
        \begin{minipage}[b]{0.45\textwidth}
            \begin{tikzpicture} 
                \begin{axis}[legend pos= south east,width=8cm,height=6.5cm,xlabel= {\# zeroes in measured pattern},ylabel={count},xmin=-0.6,xmax=4.5,ymin=20,ymax=100,title={{\bf IBM Q Melbourne}, 16 $\times$ 1024 shots},  title style={at={(1.0,0)},anchor=south east},inner axis line style={stealth-stealth}]
                \addplot+[scatter, only marks,
                    scatter src=explicit symbolic
                ] table
                [x index=0,y index=1,meta index=2,col sep=comma,header=false] {mel_zer.dat.csv};
                \addplot[dashed, mark=none, black] {64};
                \end{axis}
            \end{tikzpicture}
        \end{minipage}
        \begin{minipage}[b]{0.45\textwidth}
            \begin{tikzpicture} 
                \begin{axis}[legend pos= south west, width=8cm, height=6.5cm,xlabel= {\# zeroes in measured pattern},ylabel={count},xmin=-0.6,xmax=4.5,ymin=20,ymax=100,title={{\bf IBM Q Vigo}, 16 $\times$ 1024 shots},  title style={at={(1.0,0)},anchor=south east}]
                \addplot+[scatter, only marks,
                    scatter src=explicit symbolic
                ] table
                [x index=0,y index=1,meta index=2,col sep=comma,header=false]{vigo_zer.dat.csv};
                \addplot[dashed, mark=none, black] {64};
                \end{axis}
            \end{tikzpicture}
        \end{minipage}
        \caption{Melbourne vs. Vigo decoherence for circuits with 16 2-qubit gates}
        \label{fig:mel vs vigo}
    \end{figure*}
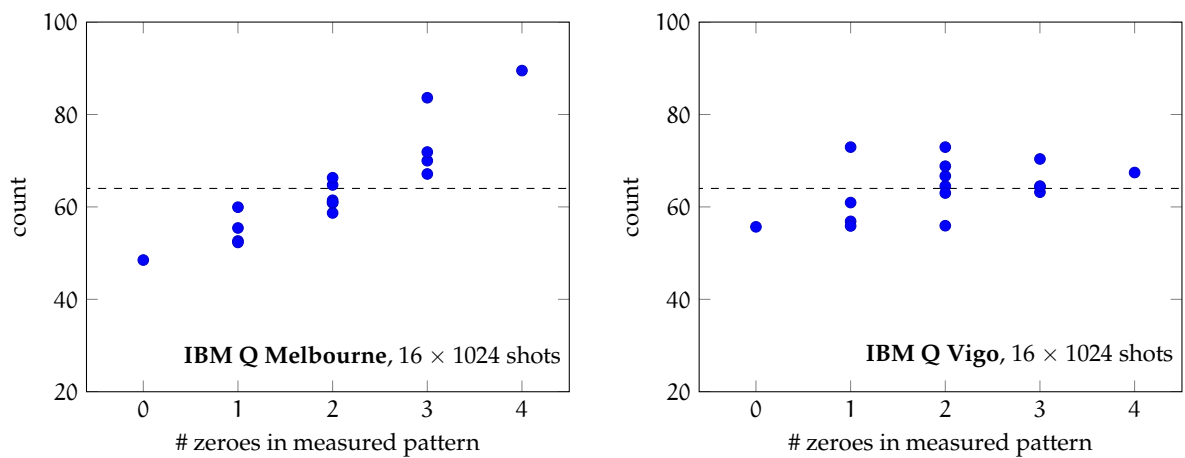 
    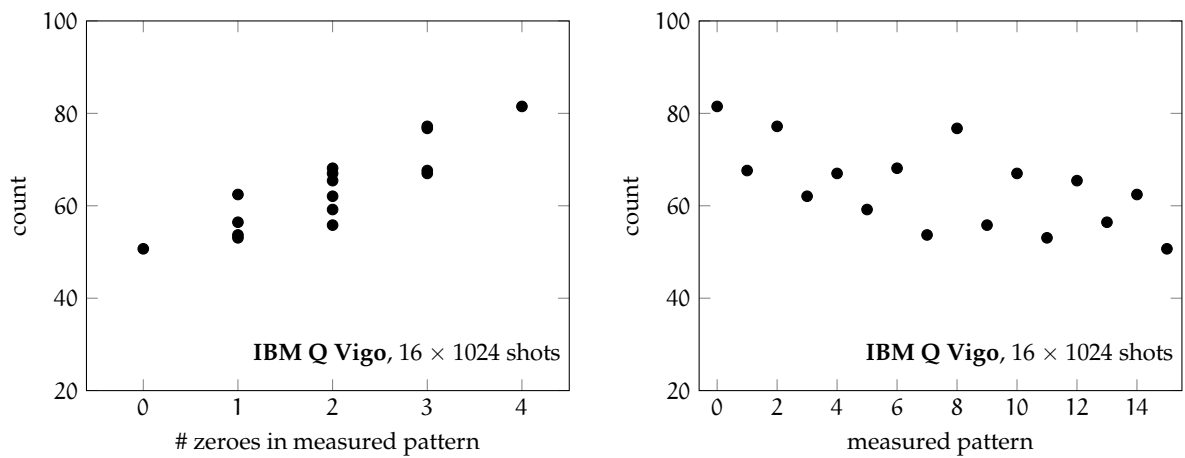
\begin{figure*}[htp]
        \begin{minipage}[b]{0.45\textwidth}
            \begin{tikzpicture} 
                \begin{axis}[legend pos= south east,width=8cm,height=6.5cm,xlabel= {\# zeroes in measured pattern},ylabel={count},xmin=-0.6,xmax=4.5,ymin=20,ymax=100,title={{\bf IBM Q Vigo}, 16 $\times$ 1024 shots},  title style={at={(1.0,0)},anchor=south east},inner axis line style={stealth-stealth}]
                \addplot [only marks] coordinates {
                    (4,81.5)
                    (3.00,	67.625)
                    (3,77.1875)
                    (2,62.0625)
                    (3,	67)
                    (2,	59.1875)
                    (2,68.125)
                    (1,	53.6875)
                    (3,	76.75)
                    (2,	55.8125)
                    (2,	67)
                    (1,	53.0625)
                    (2,	65.4375)
                    (1,	56.4375)
                    (1,	62.4375)
                    (0,	50.6875)
                };
                \end{axis}
            \end{tikzpicture}
        \end{minipage}
        \begin{minipage}[b]{0.45\textwidth}
            \begin{tikzpicture} 
                \begin{axis}[legend pos= south east,width=8cm,height=6.5cm,xlabel= {measured pattern},ylabel={count},xmin=-0.6,xmax=15.5,ymin=20,ymax=100,title={{\bf IBM Q Vigo}, 16 $\times$ 1024 shots},  title style={at={(1.0,0)},anchor=south east},inner axis line style={stealth-stealth}]
                \addplot [only marks] coordinates {
                    (0,	81.5)
                    (1,	67.625)
                    (2,	77.1875)
                    (3,	62.0625)
                    (4,	67)
                    (5,	59.1875)
                    (6,	68.125)
                    (7,	53.6875)
                    (8,	76.75)
                    (9,	55.8125)
                    (10,	67)
                    (11,	53.0625)
                    (12,	65.4375)
                    (13,	56.4375)
                    (14,	62.4375)
                    (15,	50.6875)
                };
                \end{axis}
            \end{tikzpicture}
        \end{minipage}
        \caption{IBM Q Vigo, 24 2-qubit gates}
        \label{fig:Vigo decoherence}
    \end{figure*}
    
    \clearpage
\makeatletter\onecolumngrid@pop\makeatother

%% file: main.bbl
\begin{thebibliography}{12}%
\makeatletter
\providecommand \@ifxundefined [1]{%
 \@ifx{#1\undefined}
}%
\providecommand \@ifnum [1]{%
 \ifnum #1\expandafter \@firstoftwo
 \else \expandafter \@secondoftwo
 \fi
}%
\providecommand \@ifx [1]{%
 \ifx #1\expandafter \@firstoftwo
 \else \expandafter \@secondoftwo
 \fi
}%
\providecommand \natexlab [1]{#1}%
\providecommand \enquote  [1]{``#1''}%
\providecommand \bibnamefont  [1]{#1}%
\providecommand \bibfnamefont [1]{#1}%
\providecommand \citenamefont [1]{#1}%
\providecommand \href@noop [0]{\@secondoftwo}%
\providecommand \href [0]{\begingroup \@sanitize@url \@href}%
\providecommand \@href[1]{\@@startlink{#1}\@@href}%
\providecommand \@@href[1]{\endgroup#1\@@endlink}%
\providecommand \@sanitize@url [0]{\catcode `\\12\catcode `\$12\catcode
  `\&12\catcode `\#12\catcode `\^12\catcode `\_12\catcode `\%12\relax}%
\providecommand \@@startlink[1]{}%
\providecommand \@@endlink[0]{}%
\providecommand \url  [0]{\begingroup\@sanitize@url \@url }%
\providecommand \@url [1]{\endgroup\@href {#1}{\urlprefix }}%
\providecommand \urlprefix  [0]{URL }%
\providecommand \Eprint [0]{\href }%
\providecommand \doibase [0]{http://dx.doi.org/}%
\providecommand \selectlanguage [0]{\@gobble}%
\providecommand \bibinfo  [0]{\@secondoftwo}%
\providecommand \bibfield  [0]{\@secondoftwo}%
\providecommand \translation [1]{[#1]}%
\providecommand \BibitemOpen [0]{}%
\providecommand \bibitemStop [0]{}%
\providecommand \bibitemNoStop [0]{.\EOS\space}%
\providecommand \EOS [0]{\spacefactor3000\relax}%
\providecommand \BibitemShut  [1]{\csname bibitem#1\endcsname}%
\let\auto@bib@innerbib\@empty
\bibitem [{\citenamefont {Grover}(1996)}]{grover96}%
  \BibitemOpen
  \bibfield  {author} {\bibinfo {author} {\bibfnamefont {L.~K.}\ \bibnamefont
  {Grover}},\ }in\ \href@noop {} {\emph {\bibinfo {booktitle} {Proceedings of
  the twenty-eighth annual ACM symposium on Theory of computing}}}\ (\bibinfo
  {year} {1996})\ pp.\ \bibinfo {pages} {212--219}\BibitemShut {NoStop}%
\bibitem [{\citenamefont {Grover}(2002)}]{groverfast}%
  \BibitemOpen
  \bibfield  {author} {\bibinfo {author} {\bibfnamefont {L.~K.}\ \bibnamefont
  {Grover}},\ }\href@noop {} {\bibfield  {journal} {\bibinfo  {journal}
  {Physical Review A}\ }\textbf {\bibinfo {volume} {66}},\ \bibinfo {pages}
  {052314} (\bibinfo {year} {2002})}\BibitemShut {NoStop}%
\bibitem [{\citenamefont {Arunachalam}\ and\ \citenamefont
  {de~Wolf}(2015)}]{wolfsearch}%
  \BibitemOpen
  \bibfield  {author} {\bibinfo {author} {\bibfnamefont {S.}~\bibnamefont
  {Arunachalam}}\ and\ \bibinfo {author} {\bibfnamefont {R.}~\bibnamefont
  {de~Wolf}},\ }\href@noop {} {\enquote {\bibinfo {title} {Optimizing the
  number of gates in quantum search},}\ } (\bibinfo {year} {2015}),\ \Eprint
  {http://arxiv.org/abs/1512.07550} {arXiv:1512.07550 [quant-ph]} \BibitemShut
  {NoStop}%
\bibitem [{\citenamefont {Briański}\ \emph {et~al.}(2020)\citenamefont
  {Briański}, \citenamefont {Gwinner}, \citenamefont {Hlembotskyi},
  \citenamefont {Jarnicki}, \citenamefont {Pliś},\ and\ \citenamefont
  {Szady}}]{prackum}%
  \BibitemOpen
  \bibfield  {author} {\bibinfo {author} {\bibfnamefont {M.}~\bibnamefont
  {Briański}}, \bibinfo {author} {\bibfnamefont {J.}~\bibnamefont {Gwinner}},
  \bibinfo {author} {\bibfnamefont {V.}~\bibnamefont {Hlembotskyi}}, \bibinfo
  {author} {\bibfnamefont {W.}~\bibnamefont {Jarnicki}}, \bibinfo {author}
  {\bibfnamefont {S.}~\bibnamefont {Pliś}}, \ and\ \bibinfo {author}
  {\bibfnamefont {A.}~\bibnamefont {Szady}},\ }\href@noop {} {\enquote
  {\bibinfo {title} {Introducing structure to expedite quantum search},}\ }
  (\bibinfo {year} {2020}),\ \Eprint {http://arxiv.org/abs/2006.05828}
  {arXiv:2006.05828 [quant-ph]} \BibitemShut {NoStop}%
\bibitem [{\citenamefont {Figgatt}\ \emph {et~al.}(2017)\citenamefont
  {Figgatt}, \citenamefont {Maslov}, \citenamefont {Landsman}, \citenamefont
  {Linke}, \citenamefont {Debnath},\ and\ \citenamefont {Monroe}}]{3q}%
  \BibitemOpen
  \bibfield  {author} {\bibinfo {author} {\bibfnamefont {C.}~\bibnamefont
  {Figgatt}}, \bibinfo {author} {\bibfnamefont {D.}~\bibnamefont {Maslov}},
  \bibinfo {author} {\bibfnamefont {K.}~\bibnamefont {Landsman}}, \bibinfo
  {author} {\bibfnamefont {N.~M.}\ \bibnamefont {Linke}}, \bibinfo {author}
  {\bibfnamefont {S.}~\bibnamefont {Debnath}}, \ and\ \bibinfo {author}
  {\bibfnamefont {C.}~\bibnamefont {Monroe}},\ }\href@noop {} {\bibfield
  {journal} {\bibinfo  {journal} {Nature communications}\ }\textbf {\bibinfo
  {volume} {8}},\ \bibinfo {pages} {1} (\bibinfo {year} {2017})}\BibitemShut
  {NoStop}%
\bibitem [{\citenamefont {Satoh}\ \emph {et~al.}(2020)\citenamefont {Satoh},
  \citenamefont {Ohkura},\ and\ \citenamefont {Meter}}]{kajman}%
  \BibitemOpen
  \bibfield  {author} {\bibinfo {author} {\bibfnamefont {T.}~\bibnamefont
  {Satoh}}, \bibinfo {author} {\bibfnamefont {Y.}~\bibnamefont {Ohkura}}, \
  and\ \bibinfo {author} {\bibfnamefont {R.~V.}\ \bibnamefont {Meter}},\
  }\href@noop {} {\enquote {\bibinfo {title} {Subdivided phase oracle for nisq
  search algorithms},}\ } (\bibinfo {year} {2020}),\ \Eprint
  {http://arxiv.org/abs/2001.06575} {arXiv:2001.06575 [quant-ph]} \BibitemShut
  {NoStop}%
\bibitem [{\citenamefont {Mandviwalla}\ \emph {et~al.}(2018)\citenamefont
  {Mandviwalla}, \citenamefont {Ohshiro},\ and\ \citenamefont
  {Ji}}]{Mandviwalla}%
  \BibitemOpen
  \bibfield  {author} {\bibinfo {author} {\bibfnamefont {A.}~\bibnamefont
  {Mandviwalla}}, \bibinfo {author} {\bibfnamefont {K.}~\bibnamefont
  {Ohshiro}}, \ and\ \bibinfo {author} {\bibfnamefont {B.}~\bibnamefont {Ji}},\
  }in\ \href@noop {} {\emph {\bibinfo {booktitle} {2018 IEEE International
  Conference on Big Data (Big Data)}}}\ (\bibinfo {organization} {IEEE},\
  \bibinfo {year} {2018})\ pp.\ \bibinfo {pages} {2531--2537}\BibitemShut
  {NoStop}%
\bibitem [{\citenamefont {Str{\"o}mberg}\ and\ \citenamefont
  {Blomkvist~Karlsson}(2018)}]{Stromberg}%
  \BibitemOpen
  \bibfield  {author} {\bibinfo {author} {\bibfnamefont {P.}~\bibnamefont
  {Str{\"o}mberg}}\ and\ \bibinfo {author} {\bibfnamefont {V.}~\bibnamefont
  {Blomkvist~Karlsson}},\ }\href@noop {} {\enquote {\bibinfo {title} {4-qubit
  grover's algorithm implemented for the ibmqx5 architecture},}\ } (\bibinfo
  {year} {2018})\BibitemShut {NoStop}%
\bibitem [{\citenamefont {Zhang}\ and\ \citenamefont {Korepin}(2020)}]{zhang}%
  \BibitemOpen
  \bibfield  {author} {\bibinfo {author} {\bibfnamefont {K.}~\bibnamefont
  {Zhang}}\ and\ \bibinfo {author} {\bibfnamefont {V.~E.}\ \bibnamefont
  {Korepin}},\ }\href@noop {} {\bibfield  {journal} {\bibinfo  {journal}
  {Physical Review A}\ }\textbf {\bibinfo {volume} {101}},\ \bibinfo {pages}
  {032346} (\bibinfo {year} {2020})}\BibitemShut {NoStop}%
\bibitem [{\citenamefont {Brassard}\ \emph {et~al.}(2002)\citenamefont
  {Brassard}, \citenamefont {Hoyer}, \citenamefont {Mosca},\ and\ \citenamefont
  {Tapp}}]{ampamp}%
  \BibitemOpen
  \bibfield  {author} {\bibinfo {author} {\bibfnamefont {G.}~\bibnamefont
  {Brassard}}, \bibinfo {author} {\bibfnamefont {P.}~\bibnamefont {Hoyer}},
  \bibinfo {author} {\bibfnamefont {M.}~\bibnamefont {Mosca}}, \ and\ \bibinfo
  {author} {\bibfnamefont {A.}~\bibnamefont {Tapp}},\ }\href@noop {} {\bibfield
   {journal} {\bibinfo  {journal} {Contemporary Mathematics}\ }\textbf
  {\bibinfo {volume} {305}},\ \bibinfo {pages} {53} (\bibinfo {year}
  {2002})}\BibitemShut {NoStop}%
\bibitem [{\citenamefont {Song}\ and\ \citenamefont
  {Klappenecker}(2003)}]{margolus}%
  \BibitemOpen
  \bibfield  {author} {\bibinfo {author} {\bibfnamefont {G.}~\bibnamefont
  {Song}}\ and\ \bibinfo {author} {\bibfnamefont {A.}~\bibnamefont
  {Klappenecker}},\ }\href@noop {} {\enquote {\bibinfo {title} {The simplified
  toffoli gate implementation by margolus is optimal},}\ } (\bibinfo {year}
  {2003}),\ \Eprint {http://arxiv.org/abs/quant-ph/0312225}
  {arXiv:quant-ph/0312225 [quant-ph]} \BibitemShut {NoStop}%
\bibitem [{\citenamefont {Hu}\ \emph {et~al.}(2019)\citenamefont {Hu},
  \citenamefont {Maslov}, \citenamefont {Pistoia},\ and\ \citenamefont
  {Gambetta}}]{htris}%
  \BibitemOpen
  \bibfield  {author} {\bibinfo {author} {\bibfnamefont {S.}~\bibnamefont
  {Hu}}, \bibinfo {author} {\bibfnamefont {D.}~\bibnamefont {Maslov}}, \bibinfo
  {author} {\bibfnamefont {M.}~\bibnamefont {Pistoia}}, \ and\ \bibinfo
  {author} {\bibfnamefont {J.}~\bibnamefont {Gambetta}},\ }in\ \href@noop {}
  {\emph {\bibinfo {booktitle} {Proceedings of the 56th Annual Design
  Automation Conference 2019}}}\ (\bibinfo {year} {2019})\ pp.\ \bibinfo
  {pages} {1--2}\BibitemShut {NoStop}%
\end{thebibliography}%
